 Chern Chuang <chernc@mit.edu>
 Jie Guan <guanjie@msu.edu>
 David Witalka <witalkad@msu.edu>
 Zhen Zhu <zhuzhen@msu.edu>
 Bih-Yaw Jin <byjin@ntu.edu.tw>
 David Tomanek <tomanek@pa.msu.edu>
\documentclass[aps,prb,reprint,superscriptaddress,showpacs,floatfix]{revtex4-1}
\usepackage{graphicx}% Include figure files
\usepackage{amssymb,amsmath}
\usepackage{dcolumn}% Align table columns on decimal point
\usepackage{bm}% bold math
\usepackage[mathlines]{lineno}% Enable numbering of text and display math
\begin{document}

\title{Relative Stability and Local Curvature Analysis in Carbon Nanotori }

\author{Chern Chuang}
\affiliation{Department of Chemistry,
             Massachusetts Institute of Technology,
             Cambridge, MA 02139, USA}

\author{Jie Guan}
\affiliation{Physics and Astronomy Department,
             Michigan State University,
             East Lansing, Michigan 48824, USA}

\author{David Witalka }
\affiliation{Physics and Astronomy Department,
             Michigan State University,
             East Lansing, Michigan 48824, USA}

\author{Zhen Zhu}
\affiliation{Physics and Astronomy Department,
             Michigan State University,
             East Lansing, Michigan 48824, USA}

\author{Bih-Yaw Jin}
\affiliation{Department of Chemistry and
             Center for Emerging Material and Advanced Devices,
             National Taiwan University,
             Taipei 10617, Taiwan}

\author{David Tom\'{a}nek}
\email%[E-mail: ]%
            {tomanek@pa.msu.edu}%
\affiliation{Physics and Astronomy Department,
             Michigan State University,
             East Lansing, Michigan 48824, USA}

\date{\today} % delete this line to display the current date

%---------------------------------------------------------------------
\begin{abstract}
We introduce a concise formalism to characterize nanometer-sized
tori based on carbon nanotubes and to determine their stability by
combining {\em ab initio} density functional calculations with a
continuum elasticity theory approach that requires only shape
information. We find that the high strain energy in nanotori
containing only hexagonal rings is significantly reduced in
nanotori containing also other polygons. Our approach allows to
determine local curvature and link it to local strain energy,
which is correlated with local stability and chemical reactivity.
%We present detailed theoretical analysis on the stability of
%nanometer-sized tori based on carbon nanotubes. Two classes of
%nanostructures are studied in the context of continuum elasticity:
%Tori composed of exclusively hexagonal rings and ones containing a
%minority of non-hexagonal rings. The relative contributions from
%the in-plane and out-of-plane strain to the total energy vary
%significantly in nanotori with only hexagons and the latter is
%dominant in ones with nonhexagons. Local curvature in these
%systems is shown to be correlated to the loci of the non-hexagonal
%defects, which contributes to nontrivial distribution of excess
%energy. Asymptotic analysis on three specific families of nanotori
%confirms that the current methodology is applicable across a wide
%length scale, from nanotori containing a few hundreds of atoms to
%mesoscopic structures measuring micrometers.
\end{abstract}
%---------------------------------------------------------------------


% insert suggested keywords - APS authors don't need to do this
%\maketitle must follow title, authors, abstract, \pacs, and \keywords

\maketitle
%---------------------------------------------------------------------
% \linenumbers\relax % Commence numbering lines
% If in two-column mode, this environment will change to single-column
% format so that long equations can be displayed. Use sparingly.
%\begin{widetext}
% put long equation here
%\end{widetext}

% \documentclass[12pt]{article}
%
%\onehalfspacing
%\thispagestyle{empty}

\section*{Introduction}

The past three decades witnessed an unprecedented progress in the
synthesis and use of nanostructures consisting of elemental carbon
ranging from
% advance in manipulating novel materials on the nanoscale.
% In particular, low dimensional graphitic structures are the most
%promising. They have invaluable application possibilities
% including
fullerenes to carbon nanotubes
% and nanoribbons,
and graphene\cite{{Smalley85},{Iijima91},{Novoselov04},
%{CNeto-FGuinea-KNovoselov-AGeim-review},%
%{Hirsch-review},,{awesome-allotropy}
{DT227}}.
% In addition, due
Due to the stiffness of the $\sigma$ bonding network and the
delocalization of the $\pi$ electrons in these systems, the
structure-property relationships are
% expected to be
highly nontrivial, as presence
% the effects
of local atomic defects
% can propagate on
is known to affect physical properties on the mesoscopic length
scale~\cite{Dresselhaus}. In this regard, owing to their peculiar
geometry and topology\cite{Liu-Zhao-review}, carbon nanotori based
on single-wall carbon nanotubes serve as an excellent test ground
for studying such effects.
% in studying this delocalization effect.
To name just a few, reversible elastic deformation of carbon
nanotori has been predicted\cite{Small.6.1647} and experimentally
identified\cite{JACS.133.9654}, implying potential application in
force-sensing devices. Gigantic paramagnetic response
\cite{Guo-PRL}, persistent
currents\cite{{Lin19981},{Lin19982},%
%{Latil2003},
{Rocha2004}} and even molecular anapole moments\cite{Fowler-PRL}
have been predicted for specific nanotorus geometries. Also, the
intricate shape of carbon nanotori can be exploited in the context
of host-guest chemistry and physics; for example, megahertz
oscillations of a nanotorus mounted on a
nanotube\cite{PRB.75.125415} and metal-encapsulated nanotori with
net magnetic moment\cite{PRB.76.125422} have been predicted. A
crucial prerequisite to realize these interesting physical
properties is to establish, to what degree the postulated toroidal
nanostructures are stable mechanically and thermally.
% Above all the interesting physical properties of graphitic
% structures, the thermal and mechanical stabilities which concern
% the existence of the nanostructures are the most important.

Here we study the stability of carbon nanotori formed conceptually
by bending a finite-length single-wall carbon nanotube (CNT) and
connecting its ends seamlessly. Such a nanotorus can then be
characterized the chiral index\cite{oshiyama92} $(n,m)$ of the
nanotube and number of primitive unit cells along the perimeter.
We will refer to this family of nanotori, which contain only
hexagonal rings, as to {\em polyhex} nanotori. It is known that
presence of non-hexagonal rings, such as 5-7 pairs, induces a bend
in a straight nanotube. In the following, we will refer to the
family of nanotori that contain also non-hexagonal rings as to
{\em polygonal} nanotori. We will demonstrate that a continuum
elasticity theory approach\cite{DT238}, which requires only shape
information, is capable of estimating not only the global
stability, but also local strain with a precision approaching that
of {\em ab initio} density functional calculations at a small
fraction of the computational cost.

% In this contribution we examine the stability of two classes
% of carbon nanotori. They can be easily differentiated by the
% existence of non-hexagonal rings among a majority of hexagonal
% ones. Assuming the absence of dangling bonds, a carbon nanotorus
% can be simply constructed by bending a finite straight carbon
% nanotube (CNT) and connecting its ends seamlessly. While CNTs are
% defined by the chiral vectors \cite{{oshiyama92} $(n,m)$, an
% additional positive integer specifying the number of repetitive
% units is required to characterize a carbon nanotorus. Since the
% resulting nanotori are composed of purely hexagonal rings, this
% family of nanotori is referred to as polyhex nanotori.

Our manuscript is organized as follows. In the next section we
outline the computational techniques used in our study. Next we
summarize the characteristics of different nanotori families.
Then, we analyze and obtain analytical expressions for the
in-plane and out-of-plane contributions to the elastic energy of
polyhex nanotori. In the following Section, we will present a
general overview of the elastic energy for polygonal nanotori.
Next we will discuss the distributions of curvature and elastic
energy on polygonal nanotori, and correlate them to the relative
positions of the non-hexagonal rings in the systems. This is
followed by the asymptotic analysis of three families of polygonal
nanotori with varying shape parameters, including changing
rotational symmetry, the lateral torus size normal to the axis and
the torus height along the axis.

\section*{Computational Techniques}

Our main computational approach is based on continuum elasticity
theory, which proved useful in evaluating the strain energy in
fullerenes, nanotubes and schwarzites with respect to a planar
graphene monolayer and corresponding stability
differences\cite{DT238}.

In systems with no frustration, where areas of Voronoi polygons
associated with individual atoms are at the optimum value, the
in-plane strain energy is very small. This proved to be the case
in carbon nanotubes\cite{DT238} and we expect it to hold also in
the related nanotori. Neglecting the in-plane component of strain,
the stability of nanotori could be analyzed in the inextensional
limit of the mechanical theory of a membrane\cite{Landau}, where
the elastic energy is dominated by its out-of-plane component.

The out-of-plane strain or curvature energy of a given closed
membrane can be expressed as
%by as follows.
%
\begin{equation}
\Delta E_\text{c}=D\int_{S}dA[2k^2-(1-\alpha)G]\;,%
\label{eqn:dEc}
\end{equation}
where $D$ is the flexural rigidity, $\alpha$ the Poisson ratio,
$k$ the local mean curvature, $G$ the local Gaussian curvature,
and the integration is carried out over the surface $S$ of the
membrane. If instead of the general shape we know the precise
atomic locations
% atomic detail,
from either diffraction experiments or \textit{ab initio}
calculations, we can rather use the
% is known for the molecular structure under investigation,
% the following
discretized version of Eq.~(\ref{eqn:dEc}),
% can be used instead.
%
\begin{equation}
\Delta E_\text{c} \approx A_0D \sum_i
\left[2k_i^2-(1-\alpha)G_i\right]\;.%
\label{eqn:dE_c_discrete}
\end{equation}
Here, $A_0$ is the area per atom, $k_i$ the local mean curvature
and $G_i$ the local Gaussian curvature at atom site $i$. The
summation covers all atoms of the torus. For structures comprising
of more than one element, the quantity $A_0$ may vary and needs to
be taken inside of the summation. Alongside with this expression,
accurate ways to estimate $k_i$ and $G_i$ based on the
\textit{local geometry} in the vicinity of site $i$ have been
proposed recently\cite{DT238}, which have been validated by {\em
ab initio} density functional calculations. Even though this
method relies on the knowledge of the molecular geometry, it has
been shown that even inexpensive classical force fields serve the
purpose of structure optimization surprisingly well when compared
against state-of-the-art \textit{ab initio} calculations for a
wide range of graphitic structures\cite{{Keating},{DT238}}.

The total in-plane strain energy can be estimated in the continuum
model from an integral over the entire surface area $A$,
\begin{equation}
\Delta E_s=\oint_A \epsilon(\sigma)dA\;,%
\label{eqn:Es}
\end{equation}
where $\epsilon(\sigma)$ is the energy cost per area subject to
strain $\sigma$. For specific, small strains, $\epsilon(\sigma)$
may be presented as a harmonic function of the strain $\sigma$,
with the proper coefficient taken either from experimental data or
form {\em ab initio} calculations.

Selected results based on continuum elasticity theory are
validated by {\em ab initio} calculations in the framework of
density functional theory (DFT), as implemented in the
\textsc{SIESTA} code.\cite{SIESTA} We use the local density
approximation\cite{{CA},{PZ}} to describe exchange and correlation
in the system, norm-conserving Troullier-Martins
pseudopotentials\cite{Troullier91}, and a double-$\zeta$ basis
including polarization orbitals. In the reference system, we
sample the 2D Brillouin zone of graphene by
$16{\times}16$~$k$-points.\cite{Monkhorst-Pack76} The small
Brillouin zones of isolated nanotori are sampled by only
$1$~$k$-point. We use a mesh cutoff energy of $180$~Ry to
determine the self-consistent charge density, which provides
precision in total energy of ${\alt}2$~meV/atom. All geometries
are optimized using the conjugate gradient method,\cite{CGmethod}
until none of the residual Hellmann-Feynman forces exceeds
$10^{-2}$~eV/{\AA}.

%===========< FIGURE 1 >=========================================
% Use the figure* environment if the figure should span across the
% entire page. There is no need to do explicit centering.
\begin{figure}[t]
\includegraphics[width=1.0\columnwidth]{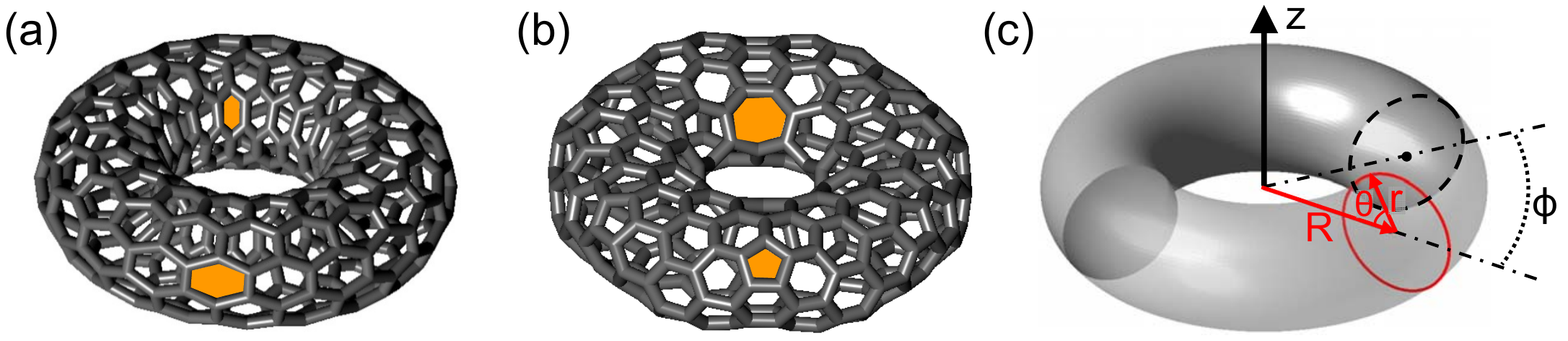}
\caption{(Color online) Atomic structure of (a) a polyhex
nanotorus with only hexagons and (b) a polygonal nanotorus
containing non-hexagonal rings. (c) A model of a perfect torus
with major radius $R$ and minor radius $r$. $\theta$ is the
longitudinal angle and $\phi$ is the azimuthal angle. The yellow
shading in (a) and (b) indicates specific polygons in the
nanotori. \label{fig:figure1}}
\end{figure}
%===========< FIGURE 1 >=========================================

\section*{Nanotorus Characteristics}

Due to the extreme flexural stiffness of straight
CNTs\cite{{DT059},{mechanical-prop-papers}}, it is expected that
the number of carbon atoms of a stable polyhex nanotorus should be
at least on the order of $10^5$. An example of a polyhex nanotorus
containing 400 carbon atoms is shown in Fig.~\ref{fig:figure1}(a).
For the two shaded hexagons located in the inner and the outer
rims of the nanotorus, it is obvious that the in-plane strain
energy is inversely proportional to the size of the nanotori. This
will be addressed in detail in the following section.

In contrast to polyhex nanotori, there has been considerable
effort in describing the second family of polygonal nanotori,
which also contain non-hexagonal rings.
% Such nanotori are referred to as polygonal nanotori since they
% are composed of a variety of polygons.
Due to the requirement of trivalency of $sp^2$ carbon atoms and
the Euler's theorem, the number of non-hexagonal rings on a
general closed surface is constrained by
\begin{equation}
\sum_{m>2}(6-m)N_m=6\chi,
\end{equation}
where $N_m$ is the number of the $m$-gonal rings and $\chi$ is the
Euler characteristic of the closed surface in question. For a
torus, we have $\chi = 0$. Thus, if we restrict ourselves to
polygonal nanotori having only pentagons, hexagons, and heptagons,
the number of pentagons ($N_5$) must equal that of heptagons
($N_7$). Apart from this simple equation, there are many more
combinatorial and geometrical constraints for a stable polygonal
carbon nanotorus to exist\cite{JCIM}. The polygonal nanotori
covered in this study are restricted to those of high symmetry
(D$_{nd}$ or D$_{nh}$ point group) and a smaller number of atoms
($<60$ atoms per rotational unit cell). One such polygonal
nanotorus is shown in Fig.~\ref{fig:figure1}(b), a representative
pentagon on the outer and a heptagon on the inner side of the
nanotorus are highlighted. As suggested before, the existence of
these non-hexagonal rings induces natural bending and thus
drastically reduces the in-plane strain energy.

\section*{Elastic Energy in Nanotori}%
\label{sec:ElasticEnergy}

As suggested above, the elastic energy within carbon nanotori can
be divided into two parts: in-plane and out-of-plane strain
energy. The in-plane strain originates from an uneven distribution
of atomic area densities as compared to graphene at equilibrium.
This is most obvious in polyhex nanotori, where hexagons along the
outer perimeter are stretched and those along the inner perimeter
are compressed, as seen in Fig.~\ref{fig:figure1}(a). The
out-of-plane strain energy is associated with forming any curved
surface of an intrinsically flat material.

In this section we examine these two contributions to the elastic
energy of nanotori. In particular, we will provide and analyze
analytical expressions for polyhex nanotori without
buckling\cite{PRB.67.195408}.
% approximated by the parametrization in Eq.~(\ref{eqn:para}),
% are obtained and analyzed.
For polygonal nanotori, we will show that the contribution from
the in-plane strain is negligible, similar to the case of small
fullerenes\cite{DT238}.

\subsection*{In-Plane Strain Energy in Polyhex Nanotori}

Points on the surface of a perfect parametric torus are described
by
% given by the following expression.
%
\begin{eqnarray}
\begin{cases}
x=(R-r\cos\theta)\cos\phi\;,\\
y=(R-r\cos\theta)\sin\phi\;,\\
z=r\sin\theta\;.\end{cases}%
\label{eqn:para}
\end{eqnarray}

%===========< FIGURE 2 >========================================
\begin{figure}[t]
\includegraphics[width=0.7\columnwidth]{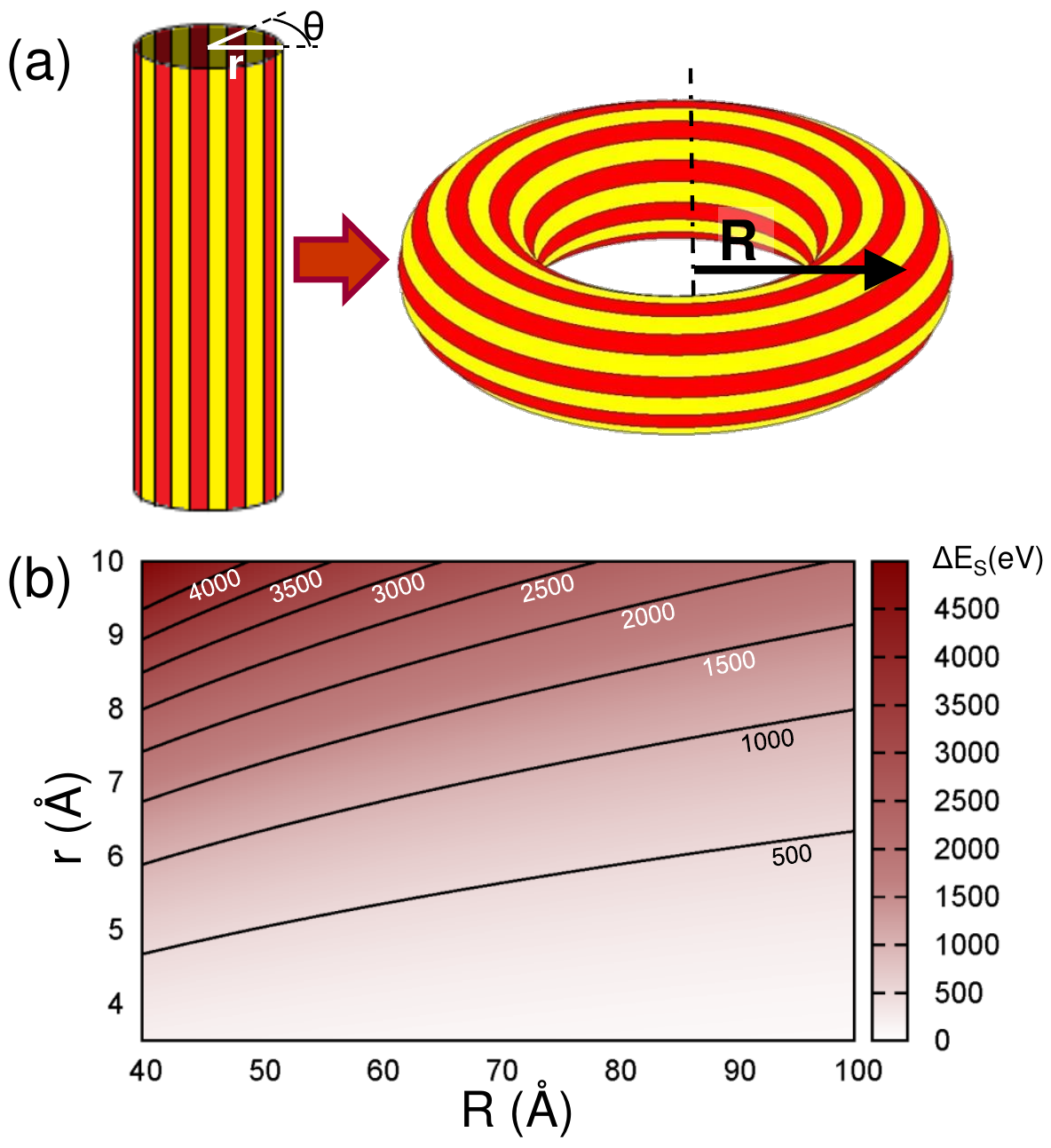}
\caption{(Color online) (a) Schematic of a striped nanotube
(left), which is bent to a nanotorus (right). (b) Contour plot
representing the in-plane strain energy ${\Delta}E_s(R,r)$
according to Eq.~(\ref{eqn:Es2}).
%and (c) out-of-plane strain energy ${\Delta}E_C$ given in
%Eq.\ref{eqn:dEc_torus} of nanotori with different $R$ and $r$.
\label{fig:figure2} }
\end{figure}
%===========< FIGURE 2 >========================================

Here and in Fig.~\ref{fig:figure1}(c), $R$ is the major and $r$
the minor radius, $\theta$ the zenith angle, and $\phi$ the
azimuthal angle. Note that $R>r$ is required for a normal ring
torus without self-intersection. A polyhex nanotorus, shown in
Fig.~\ref{fig:figure1}(a), can be constructed by rolling up a
finite CNT and connecting the two ends together, as indicated in
Fig.~\ref{fig:figure2}(a). The major radius $R$ of the polyhex
nanotorus can be approximated by the length of the original
nanotube divided by $2\pi$. In this case, the area density of
atoms is smaller than the average along the outer perimeter of the
torus, characterized by $\pi/2<\theta<3\pi/2$, and larger along
the inner perimeter. Obviously, the in-plane strain will be
negligible in comparison to the out-of-plane strain if $R>>r$.
Should this not be the case, then the in-plane strain energy will
be important, as it describes the significant distortion of
hexagonal rings along the inner and the outer perimeters of a
polyhex nanotorus, which are highlighted in
Fig.~\ref{fig:figure1}(a).

Here we present a simple estimation of the in-plane strain energy
for polyhex nanotori without buckling. As shown in
Fig.~\ref{fig:figure2}(a), a CNT can be seen as parallel narrow
strips or nanoribbons of equal length $2{\pi}R$ that are
% (polyacene) -- only for armchair
connected side-by-side. After the CNT is deformed to a nanotorus,
the strips on the inside of the nanotorus are compressed and the
strips on the outside are stretched. We further assume that the
contributions from all other in-plane distortion modes are
negligible. As shown in
% Fig.~\ref{fig:figure1}(a) and
Fig.~\ref{fig:figure2}(a), the width of the strips is ${r}d\theta$
and the length is $2{\pi}(R-r\cos\theta)$. The strain $\sigma$ of
the strips can be written as
\begin{equation}
\sigma (\theta)=\frac{2{\pi}(R-r\cos\theta)-2{\pi}R}{2{\pi}R} =
-\frac{r}{R}\cos\theta\;.%
\label{eqn:sigma}
\end{equation}
The total in-plane strain energy of the nanotorus is just the sum
of the strain energies of all the strips,
\begin{equation}
\Delta E_s=\int_{0}^{2\pi} \epsilon(\sigma)
2\pi(R-r\cos\theta)rd\theta\;.%
\label{eqn:Es}
\end{equation}
Here, $\epsilon(\sigma)$ is the energy cost per area of graphene
subject to uniaxial strain $\sigma$. For sufficiently small strain
($\sigma{\lesssim}5\%$), $\epsilon(\sigma)$ is a parabolic
function of $\sigma$,
\begin{equation}
\epsilon(\sigma)=c\sigma(\theta)^2=
c\frac{r^2}{R^2}\cos^2\theta\;.%
\label{eqn:epsilon}
\end{equation}
A fit to graphene data yields the numerical value
$c=9.94$~eV/{\AA}$^2$. Substituting Eq.~(\ref{eqn:epsilon}) into
Eq.~(\ref{eqn:Es}), we obtain an expression for the total in-plane
strain energy of a nanotorus
\begin{eqnarray}
\nonumber %
\Delta E_s&=&\int_{0}^{2\pi}c\frac{r^2}{R^2}
            \cos^2\theta 2\pi(R-r\cos\theta)rd\theta\\
          &=&2\pi^2c\frac{r^3}{R}\;.
\label{eqn:Es2}
\end{eqnarray}
The behavior of ${\Delta}E_s(R,r)$ in polyhex nanotori is shown in
Fig.~\ref{fig:figure2}(b). As suggested by Eq.~(\ref{eqn:Es2}),
the strain energy is proportional to $r^3$ and inversely
proportional to $R$. Note that this expression is obtained in the
harmonic limit of small strain, which translates into $R>>r$.
Also, we have assumed that the nanotori are ideal and described by
Eq.~(\ref{eqn:para}). In reality, the cross-section of an elastic
nanotorus may deviate from a perfect circle upon relaxation, and
buckling may occur along the perimeter. In spite of its limits,
this expression provides a concrete estimate of the strain energy
of experimentally observed rings based on CNTs%
\cite{{reference-20-a},{reference-20-b},%
{reference-20-c},{reference-20-d}}. In the reported cases, $R$ is
of the order of microns and $r$ of the order of nanometers, so all
assumptions in deriving Eq.~(\ref{eqn:Es2}) are justified.

%===========< FIGURE 3 >=========================================
\begin{figure*}[t]
\includegraphics[width=1.6\columnwidth]{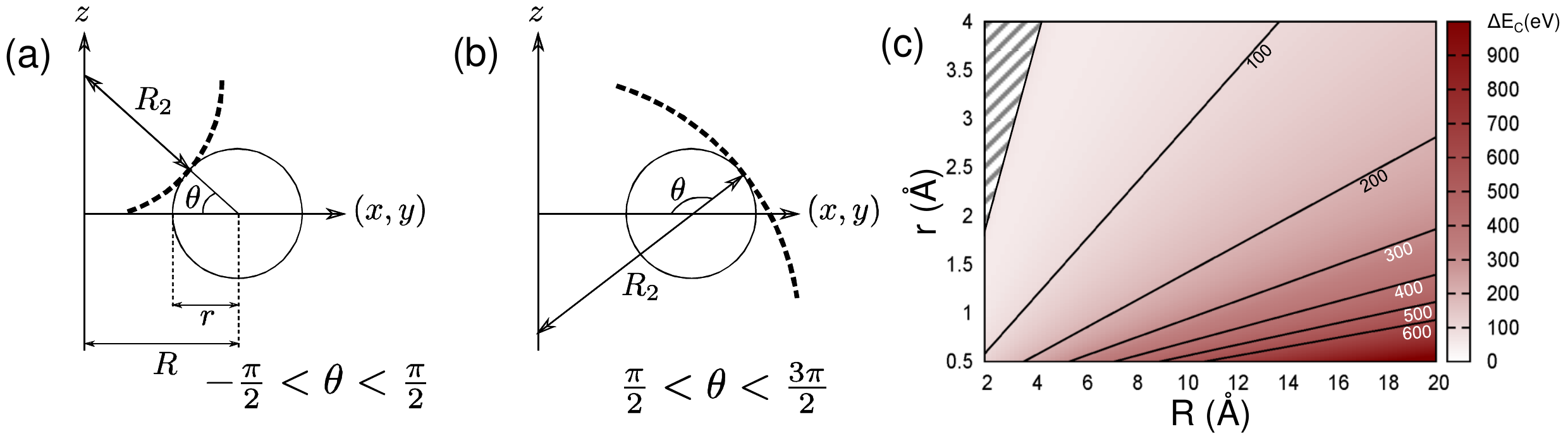}
\caption{(Color online) Definition of the principal radius $R_2$
of a nanotorus in (a) its interior and (b) its exterior part as a
function of the zenith angle $\theta$. (c) Contour plot displaying
the out-of-plane curvature energy ${\Delta}E_c(R,r)$ of nanotori
according to Eq.~(\ref{eqn:dEc_torus}).
%Part (c) shows the curvature energy of
%a perfect parametric torus as given in Eq.\ref{eqn:dEc_torus}.
\label{fig:figure3}}
\end{figure*}
%===========< FIGURE 3 >========================================

\subsection*{Out-of-Plane Strain Energy in a Perfect Parametric Torus}

Besides the in-plane strain, there is also an out-of-plane strain
caused by deviation from planarity, which is represented by
Eq.~(\ref{eqn:dEc}). For the parametric torus given in
Eq.~(\ref{eqn:para}), we infer the infinitesimal area $dA$ in
Eq.~(\ref{eqn:dEc}) to be $dA=rd\theta\cdot(R-r\cos\theta)d\phi$.
Obviously $r$ is one of the principal radii of curvature, which is
arbitrarily chosen to be $R_1$. The situation is more complicated
for $R_2$, the other principal radius of curvature. As illustrated
in Fig.~\ref{fig:figure3}, $R_2$ is given by
% the following expression
%
\begin{eqnarray}
R_2(\theta)=
\begin{cases}
R\sec\theta-r\quad&{\rm{for}}\;-\frac{\pi}{2}<\theta<\frac{\pi}{2}\\
r-R\sec\theta\quad&{\rm{for}}\;+\frac{\pi}{2}<\theta<\frac{3\pi}{2}\;.
\end{cases}
\end{eqnarray}
Clearly, $R_2\rightarrow\infty$ when $\theta=\pm\frac{\pi}{2}$,
which is indeed the case since the torus is tangent to the planes
$z=\pm{r}$.

Now the evaluation of ${\Delta}E_c$ according to
Eq.~(\ref{eqn:dEc}) is straightforward. Since the integrand is
independent of $\phi$, we end up with the expression
\begin{eqnarray}
\Delta E_c(R,r)&=&
\pi Dr\int_0^{2\pi}d\theta(R-r\cos\theta)\times\nonumber\\
& &\left(\frac{1}{R_1^2}+%
\frac{1}{R_2(\theta)^2}+\frac{2\alpha}{R_1R_2(\theta)}\right)
\nonumber\\
&=&2\pi^2D\frac{R^2}{r\sqrt{(R+r)(R-r)}}\;.%
\label{eqn:dEc_torus}
\end{eqnarray}
In contrast to perfectly spherical fullerenes, where $\Delta E_c$
is independent of radius\cite{DT238}, the curvature energy for
nanotori depends both on the major radius $R$ and the minor radius
$r$. The out-of-plane curvature energy ${\Delta}E_c(R,r)$ of
carbon nanotori is presented in Fig.~\ref{fig:figure3}(c).

We find that the curvature energy does not depend on the Poisson
ratio $\alpha$, a result from the Gauss-Bonnet theorem which
states the Gaussian curvature integrated over a closed surface
equals the Euler characteristic ($\chi$) of the surface times
$2\pi$. Since $\chi=0$ for nanotori or any tubular structures,
${\Delta}E_c$ does not depend on $\alpha$. Note that the
$1/\sqrt{R-r}$ term in the expression diverges at $R=r$,
corresponding to a horn torus, and becomes imaginary for $R<r$,
representing spindle nanotori with self-intersection in the
center. In addition, taking the limit of $R\rightarrow\infty$,
this expression reduces to the curvature energy of a straight
cylinder of length $l=2\pi R$,
\begin{equation}
\Delta E_c=\frac{\pi D l}{r}\;.%
\label{eqn:dEc_CNT}
\end{equation}
%
% where $l=2\pi R$ is the length of the cylinder.
Also, by setting the derivative of $\Delta E_c$ with respect to
$R$ to zero, one obtains the minimal torus curvature energy of
${\Delta}\bar{E_c}=4\pi^2D$ for $R=\sqrt{2}r$, as also reported
previously\cite{{reference-21-a},{reference-21-b}}.

As mentioned before, for polyhex nanotori described by
Eq.~(\ref{eqn:para}) and $R>>r$, the in-plane strain energy can be
adequately quantified by Eq.~(\ref{eqn:Es2}) and the out-of-plane
strain energy by Eq.~(\ref{eqn:dEc_CNT}). Taking polyhex nanotori
constructed from $(10,10)$ CNTs with $r=7$~{\AA} as an example,
the constraint on the major radius $R$ for the harmonic
approximation in Eq.~(\ref{eqn:epsilon}) to hold is estimated to
be $R{\agt}150$~{\AA}. For $R=150$~{\AA} we find that the in-plane
strain energy ${\Delta}E_s=450$~eV and the out-of-plane strain
energy ${\Delta}E_c=600$~eV are of the same order. On the other
hand, the majority of rings \cite{JPCB.103.7551} synthesized by
ultrasonic wave treatment of CNTs in solution has $R=3500$~{\AA}.
In those rings, the contribution from the in-plane strain,
${\Delta}E_s=20$~eV, is negligible in comparison with the
out-of-plane strain of ${\Delta}E_c=1.4{\times}10^4$~eV.

%===========< FIGURE 4 >========================================
\begin{figure}[b]
\includegraphics[width=1.0\columnwidth]{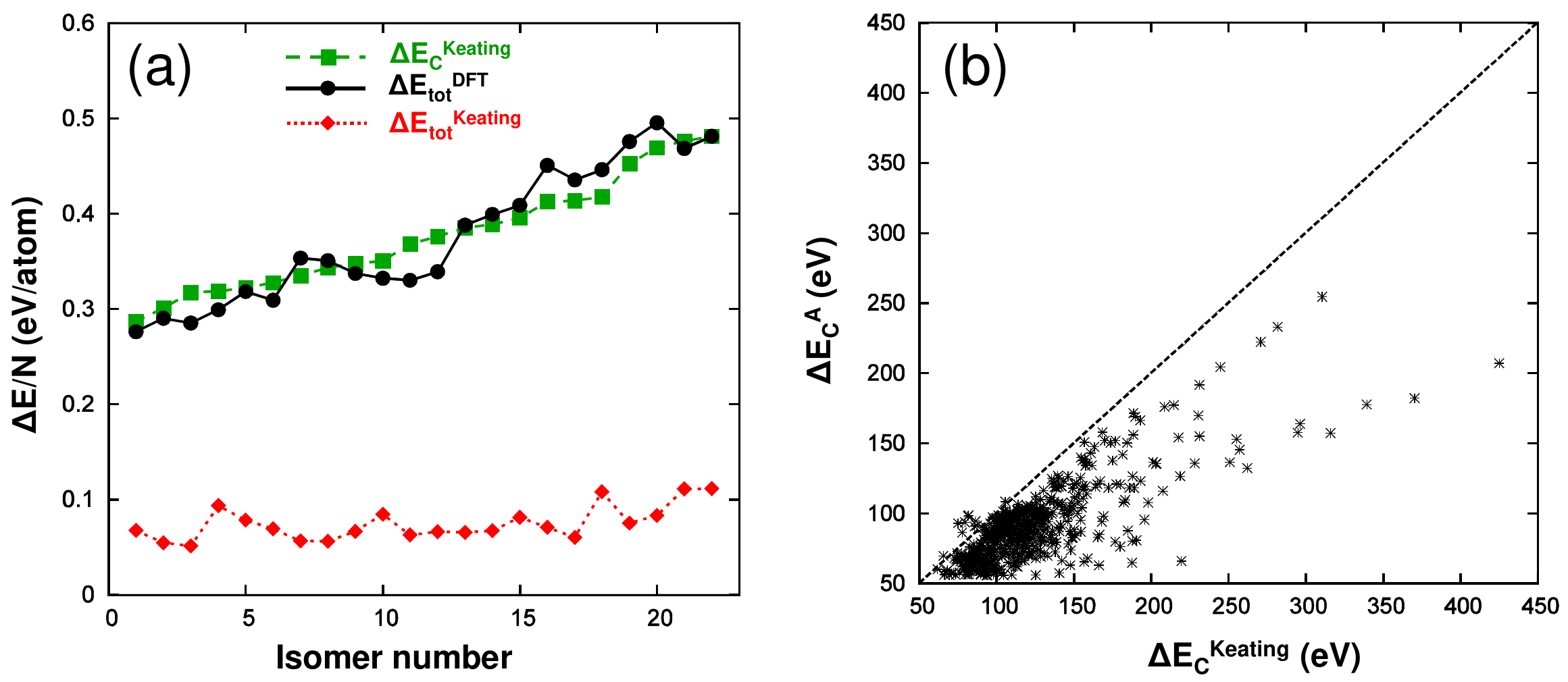}
\caption{(Color online) Strain energy ${\Delta}E$ in different
polygonal nanotori. (a) Energy differences per atom
${\Delta}E_{tot}^{DFT}$ based on DFT and
${\Delta}E_{tot}^{Keating}$ based on the Keating potential are
presented next to curvature energies ${\Delta}E_c^{Keating}$ based
on Eq.~(\ref{eqn:dE_c_discrete}) for Keating-optimized geometries.
See the main text for detailed description. (b) Comparison between
the analytical expression in Eq.~(\ref{eqn:dEc_torus}) for the
curvature energy ${\Delta}E_c^A$ based on the elastic description
of a parametric torus and the curvature energy
${\Delta}E_c^{Keating}$ obtained using
Eq.~(\ref{eqn:dE_c_discrete}) for Keating-optimized discrete
torus geometries.%
\label{fig:figure4} }
\end{figure}
%===========< FIGURE 4 >========================================

\subsection*{Strain Energy in Polygonal Nanotori}

In general, the shape of a polygonal nanotorus deviates from that
of a perfect parametric torus due to the presence of non-hexagonal
defects in a hexagonal network. The largest changes in the
curvature occur near individual non-hexagonal defects, and often
there are relatively flat segments in-between the defects.
Consequently, the analytical expression in
Eq.~(\ref{eqn:dEc_torus}) should only be used as a very
approximate way to estimate the strain energy of a polygonal
nanotorus that is roughly characterized by a set of effective
torus radii $(R,r)$. For a quantitatively better energy estimate,
we must use the discretized version of Eq.~(\ref{eqn:dEc}),
Eq.~(\ref{eqn:dE_c_discrete}), that takes into account the
specific shape of a carbon nanotorus. As mentioned previously, we
first obtain the molecular geometry from the classical Keating
force field\cite{Keating}.

Once the optimum geometry is established, the local mean curvature
$k_i$ and the local Gaussian curvature $G_i$ are determined
everywhere and substituted into Eq.~(\ref{eqn:dE_c_discrete}). The
results of this methodology to a subset of 22 polygonal nanotori
are shown in Fig.~\ref{fig:figure4}(a). Here the strain energy
calculated through Eq.~(\ref{eqn:dE_c_discrete}) with geometry
optimized by Keating potential is represented by the green
squares, energy calculated with the accurate DFT method by black
dots, and the Keating potential energy for the Keating-optimized
optimized geometry by red rhombi. Our results show clearly that
for Keating-optimized geometries, strain energies based on
Eq.~(\ref{eqn:dE_c_discrete}) reproduce our {\em ab initio}
results rather well. On the other side, strain energies estimated
using Keating potential alone not only significantly
underestimates the strain, but also do not follow the correct
general trend. This firmly establishes the applicability of the
continuum methodology to polygonal nanotori under investigation.

For specific nanotori, local geometric features that are not
described by Eq.~(\ref{eqn:para}) contribute to strain energy in
addition to Eq.~(\ref{eqn:dEc_torus}). Corresponding results are
presented
% This is studied
in Fig.~\ref{fig:figure4}(b). We considered a large set of
polygonal nanotori, which were optimized by the Keating potential,
and fitted a pair of torus radii $(R,r)$ for each of them. We then
correlated the strain energy ${\Delta}E_c^A$ obtained using the
analytical expression in Eq.~(\ref{eqn:dEc_torus}) with the more
proper value ${\Delta}E_c^{Keating}$ based on the optimum discrete
geometry, based on Eq.~(\ref{eqn:dE_c_discrete}). As expected,
most data points lie below the dashed
${\Delta}E_c^A={\Delta}E_c^{Keating}$ line. Even though the strain
energies estimated using the two approaches appear proportional to
each other, it is clear that the analytical expression in
Eq.~(\ref{eqn:dEc_torus}) underestimates strain in polygonal
nanotori significantly. We conclude that while a quick estimate of
the strain energy based on Eq.~(\ref{eqn:dEc_torus}) is useful
once the global parameters $(R,r)$ are known,
Eq.~(\ref{eqn:dE_c_discrete}) should be used if quantitative
comparison among different structures is intended.

\section*{Local Stability Analysis of Polygonal Nanotori}
\label{sec:LocalStability}

In the previous Section, we have considered nanotori as whole
objects and established elasticity theory as a valid tool to
estimate strain energy with respect to planar graphene in
different torus isomers. The total strain energy ${\Delta}E$,
according to Eqs.~(\ref{eqn:dEc}),(\ref{eqn:dE_c_discrete}),
(\ref{eqn:Es}), may be represented as an integral or a sum of
local contributions. In other words, unlike in more complex DFT
calculations, this approach allows to estimate local contributions
towards the total strain energy. In the following, we will
investigate the local strain distributions within individual
nanotori. Next, we will apply this approach to analyze three
series of polygonal nanotori and discuss trends in the strain
energy in the light of a corresponding asymptotic analysis.

% While the previous section corroborates the foundation of applying
% elasticity theory to polygonal nanotori, all benchmark and
% comparisons are made among different nanotori. On the other hand,
% the power of this methodology essentially lies on the fact that
% the contribution of local sites to the total energy can be
% quantified as the summand of Eq.~(\ref{eqn:dE_c_discrete}). Here
% we look into further details of the local distributions within
% individual nanotori. And this methodology is further applied to
% analyze three series of polygonal nanotori and compared to their
% respective asymptotic analysis.

%===========< FIGURE 5 >========================================
\begin{figure}[!tb]
\includegraphics[width=1.0\columnwidth]{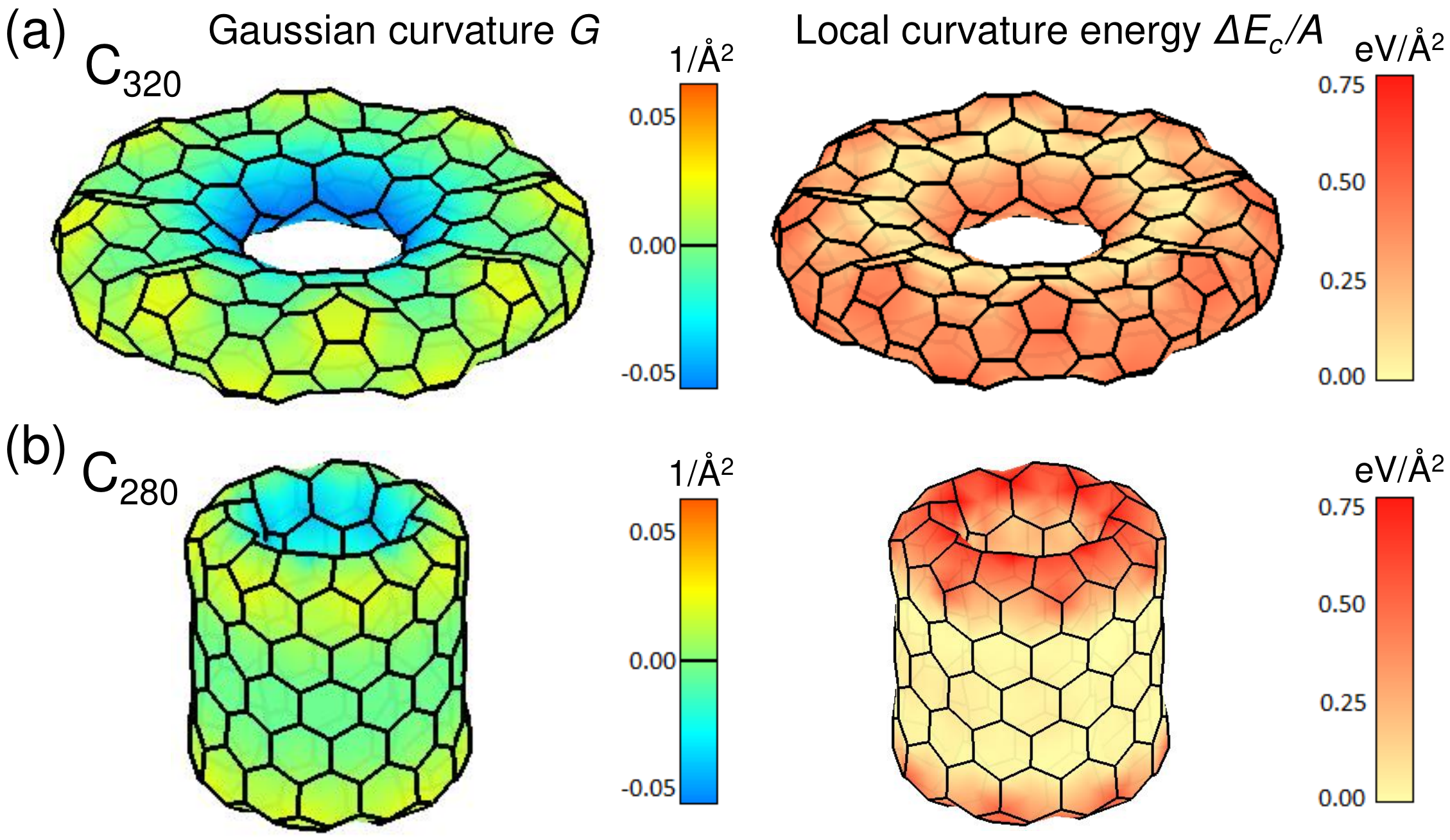}
\caption{(Color online)%
Local Gaussian curvature $G$ (left panels) and local curvature
energy ${\Delta}E_c/A$ (right panels) across the surface of
selected nanotori. Representative examples shown are (a) the
flattened C$_{320}$ nanotorus with 320 atoms and (b) the elongated
C$_{280}$ nanotorus with 280 atoms. $G$ and ${\Delta}E_c/A$ are
interpolated from their values at the atomic sites.
\label{fig:figure5} }
\end{figure}
%===========< FIGURE 5 >=========================================

\subsection*{Local Curvature Distribution}

The local geometry of a smooth, two-dimensional object may be
characterized by two independent quantities, such as the local
mean curvature $k$ and the local Gaussian curvature $G$. These
quantities can be used in Eq.~(\ref{eqn:dEc}) or its discretized
counterpart, Eq.~(\ref{eqn:dE_c_discrete}), to determine the
curvature energy. The distribution of the local Gaussian curvature
$G_i$ and the local curvature energy per area,
${\Delta}E_c^{(i)}/A_0=D[2k_i^2-(1-\alpha)G_i]$, across the
surface of two representative nanotori is shown in
Fig.~\ref{fig:figure5}.
% we present a different set, the Gaussian
% curvature (left panel) and the local curvature energy per area
% ($\Delta E_c^{(i)}/A_0=D[2k_i^2-(1-\alpha)G_i]$, right panel), for
% two selected polygonal nanotori with different geometric features.

In both cases, and others shown in the Supplemental
Material\cite{torus15-SM}, the positively curved segments, shown
in red in the left panels, are concentrated near the loci of
pentagons along the outer perimeter. The negatively curved
segments, shown in blue, are concentrated near the loci of
heptagons along the inner perimeter. Specifically, the heptagons
along the inner perimeter of the C$_{320}$ nanotorus in
Fig.~\ref{fig:figure5}(a) are well separated from the pentagons
along the outer perimeter. This is different from the axially
elongated C$_{280}$ torus of Fig.~\ref{fig:figure5}(b), where
pentagon-heptagon pairs a the upper and lower planes form an
azulene-like pattern. As a consequence, the Gaussian curvature is
more evenly distributed in the latter. We emphasize again that for
closed nanotori, the summation of the local Gaussian curvatures is
strictly zero as dictated by the Gauss-Bonnet theorem.

The distribution of the local curvature energy is even more
intriguing. Within many nanotori we investigated, some of which
are discussed in the Supplemental Material\cite{torus15-SM}, we
observed some degree of correlation between the absolute value of
the local Gaussian curvature and the curvature energy. This should
imply, at least for the nanotori investigated, that the two local
curvatures $k_i$ and $G_i$ are not entirely independent. Yet the
worth of this correlation has its limitations, as shown in
Fig.~\ref{fig:figure5}. Whereas in the flattened C$_{320}$
nanotorus in Fig.~\ref{fig:figure5}(a) the curvature energy is
rather evenly distributed across the structure, the strain is
clearly largest near the upper and lower ends of the C$_{280}$
nanotorus in Fig.~\ref{fig:figure5}(b). This curvature energy
distribution in the right panels differs obviously from the
Gaussian curvature distribution in the left panels. The reason for
this finding is that in these extreme structures, we can not truly
decouple $k_i$ and $G_i$. In C$_{280}$, $G_i{\approx}0$ and $k_i$
is constant in the central `tubular segments'. Only at the upper
and lower ends, a large mean curvature $k_i$ is required to
connect the inner and the outer tube. The flatter C$_{320}$
nanotorus lacks `tubular segments' with $G_i{\approx}0$.
Therefore, the Gaussian curvature and curvature energy are better
correlated and more evenly distributed in this isomer. The
different shapes of carbon nanotori will be discussed in more
detail later on.

We need to point out that the determination of the local Gaussian
curvature $G_i$ and the local mean curvature $k_i$ is a
non-trivial task that requires extra attention in discrete,
irregular structures. According to the procedure outlined in
Ref.~\onlinecite{DT238}, the determination of $G_i$ requires
second-nearest-neighbor information. On the other hand, the
trivalency of graphitic carbon system lends itself to a compact
definition of $k_i$ based on nearest neighbors only. The asymmetry
in handling the two curvatures leads to the possibility of
negative local curvature energies according to
Eq.~(\ref{eqn:dE_c_discrete}), which is unphysical. This can be
resolved by lowering the resolution of $k_i$, e.g. averaging its
values using a window that also includes second-nearest neighbors.

%===========< FIGURE 6 >=========================================
\begin{figure}[t]
\includegraphics[width=1.0\columnwidth]{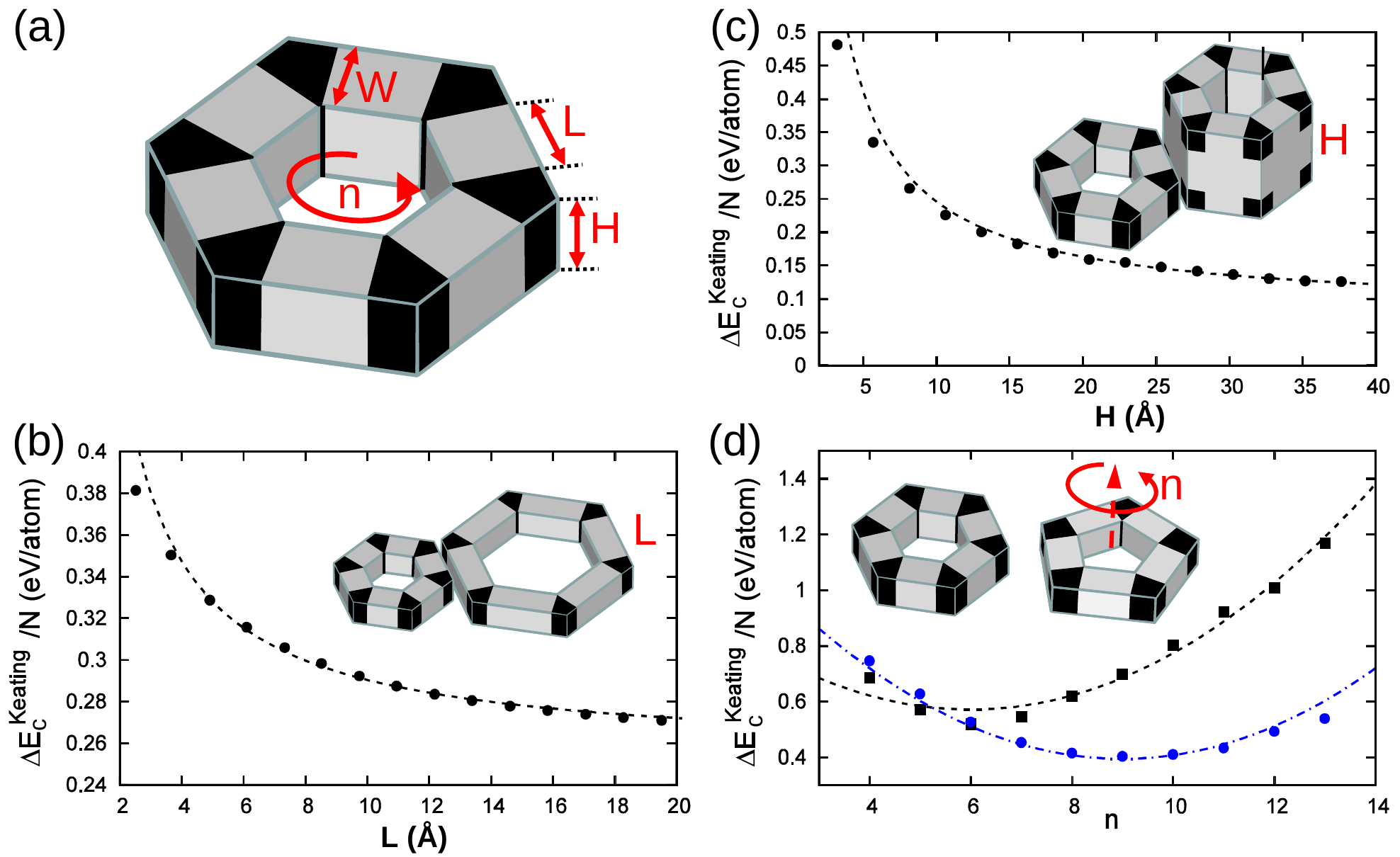}
\caption{(Color online) (a) A schematic model of a polygonal
nanotorus with $n$-fold symmetry ($n=6$ is shown here), consisting
of $n$ nanotube segments that are connected by $n$ nanotube elbow
joints. The nanotube segments are characterized by the width $W$,
length $L$ and height $H$. (b-d) Curvature energies per atom based
on Eq.~(\ref{eqn:dE_c_discrete}) for Keating-optimized nanotori of
different shapes. ${\Delta}E_c$ is presented for nanotori with (b)
$n, W, H = const.$, but nanotube length $L$ changing, (c) $n, L, W
= const.$, but nanotube height $H$ changing, and (d) $L, W, H =
const.$, but rotational symmetry number $n$ changing. The insets
show nanotorus models with different values of $L$, $H$ and $n$.
The lines in (b) and (c) are analytical extrapolations discussed
in the text. Two sets of data points in (d), connected by lines to
guide the eye, correspond to two structural families of nanotori,
described in the text. \label{fig:figure6}}
\end{figure}
%===========< FIGURE 6 >==========================================

\subsection*{Shape Dependence of Stability}

In spite of a large number of published reports%
\cite{{reference-29-a},{reference-29-b},{reference-21-a},%
{reference-21-b},{reference-31-32-a},{reference-31-32-b},%
{reference-31-32-c},{reference-31-32-d},{reference-31-32-e},%
{reference-33-a},{reference-33-b}}, a systematic investigation of
how the nanotorus stability depends on its shape has been missing.
Here we intend to cover this gap by examining the dependence of
the total curvature energy on main parameters defining the shape
of polygonal carbon nanotori, namely the rotational symmetry
number $n$, the length $L$ and the width $W$ of the nanotorus
segments, and the height $H$ of the nanotorus, shown schematically
in Fig.~\ref{fig:figure6}(a). A more detailed classification
scheme of polygonal nanotori is provided in References
\onlinecite{{reference-34-b},{JCIM}}. In the following, we discuss
the dependence of the curvature energy per atom ${\Delta}E_c$ on
the shape parameters. The structures used to obtain the numerical
results in Fig.~\ref{fig:figure6}(b-d) are displayed in the
Supplemental Material\cite{torus15-SM}, along with the
distribution of local Gaussian curvature and curvature energy
following the convention used in Fig.~\ref{fig:figure5}.

\subsubsection*{Side Length Dependence}

We first look at an interesting case of a polygonal nanotorus
consisting of $n$ straight nanotube segments of finite length $L$
that are connected by $n$ elbow joints as illustrated in
Fig.~\ref{fig:figure6}(a). As suggested in the insets of
Fig.~\ref{fig:figure6}(b), increasing the length of the nanotube
segments reduces gradually the influence of the joints and for
large $L$, the energy of the elbow joints becomes negligibly small
when compared to the curvature energy of the straight tubular
segments. This is illustrated for a specific family of nanotori in
Fig.~\ref{fig:figure6}(b), where the data points represent results
obtained using Eq.~(\ref{eqn:dE_c_discrete}) for a set of
Keating-optimized nanotori with different side length $L$ that are
displayed in the Supplemental Material\cite{torus15-SM}.

For this particular series of nanotori, we may express the
curvature energy per atom by ${\Delta}E_\text{c}(L)/N=a+b/L$ and
display this dependence, with $a=0.254$~eV and
$b=0.357$~eV$\cdot${\AA}, by the dashed line in
Fig.~\ref{fig:figure6}(b).

A more specific expression can be derived assuming that the
nanotube segments are characterized by the chiral index $(m,n)$.
In that case, the nanotube radius
$r=1.42$~{\AA}${\times}\sqrt{m^2+mn+n^2}{\times}\sqrt{3}/(2\pi)$
can be used in Eq.~(\ref{eqn:dEc_CNT}). For a nanotube segment of
length $L$, the number of atoms can be estimated using
$N=2{\pi}rL/A_0$, where
$A_0=(1.42$~{\AA}$)^2{\times}3\sqrt{3}/4=2.62$~{\AA$^2$} is the
area per atom in graphene. Then, we obtain
\begin{equation}
\frac{{\Delta}E_c(m,n)}{N} = %
A_0\frac{{\Delta}E_c}{2{\pi}rL} =%
\frac{\sqrt{3}\pi^2D}{2(m^2+mn+n^2)}\;.%
\label{eqn:dEc_cvec}
\end{equation}
%
% where $A_0=\sqrt{3}a_0^2/4$ the area per atom has been inserted.
The nanotori in Fig.~\ref{fig:figure6}(b) are based on $(4,4)$
CNTs, yielding ${\Delta}E_c/N=0.251$~eV, in very good agreement
with the fitted constant $a$. The residual energy term $b$
represents the local contribution from the elbow joints, which
becomes negligibly small in the large $L$ limit.

\subsubsection*{Height Dependence}

Next, we study a series of polygonal nanotori resembling
double-walled CNTs with their adjacent wall ends connected by a
lip-lip interaction\cite{DT227} in the form of a graphitic
network, shown in Fig.~\ref{fig:figure6}(c). With increasing
height $H$ of the nanotorus, the curvature energy per atom
${\Delta}E_c/N$ will be increasingly dominated by the central
tubular part. In analogy to arguments used for nanotori with
varying $L$, the total number of atoms $N$ is proportional to $H$
except a finite number at the double-wall nanotube ends. Then, the
curvature energy per atom can be expressed by
${\Delta}E_c(H)/N=a'+b'/H$. This behavior, with the values
$a'=0.080$~eV and $b'=1.658$~eV$\cdot${\AA}, is reproduced by the
dashed line in Fig.~\ref{fig:figure6}(c) for a particular series
of nanotori depicted in the Supplemental
Material\cite{torus15-SM}.

Similar to the case discussed in Fig.~\ref{fig:figure6}(b),
${\Delta}E_c/N$ can be approximated by the sum of curvature
energies of two nanotubes of length $H$, the inner one with radius
$r_i$ and the outer one with radius $r_o$. Then,
% Also, similar to the analysis presented previously, the energy per
% atom as a function of chiral vectors $\vec{c}_i=(c_{i,1},c_{i,2})$
% (inner tube) and $\vec{c}_o=(c_{o,1},c_{o,2})$ (outer tube) can be
% derived.
%
\begin{eqnarray}
\frac{{\Delta}E_c(r_i,r_o)}{N} = %
A_0\frac{{\Delta}E_c(r_i)+{\Delta}E_c(r_o)}{2\pi(r_i+r_o)H}= %
\frac{{A_0}D}{2{r_i}{r_o}} \\ %
= \frac{\sqrt{3}{\pi^2}D}
{2\sqrt{(m_i^2+m_in_i+n_i^2)(m_o^2+m_on_o+n_o^2)}}\;. \nonumber %
\end{eqnarray}
For the nanotori presented in Fig.~\ref{fig:figure6}(c), the
chiral indices are $(m_i,n_i)=(5,5)$ and $(m_o,n_o)=(10,10)$,
which gives the value ${\Delta}E_c/N=0.080$~eV. This value, again,
is in perfect agreement with the fitted value $a'$ above. Again,
the constant $b'$ term describes the residual energy of the end
joints connecting the inner and the outer tube.

In the above energy estimate, we have ignored the inter-layer
interaction between the outer and inner wall. If the separation
$r_o-r_i$ between the walls were as small as in graphite, this
stabilizing interaction would reduce the strain energy by
${\approx}0.03$~eV/atom. In reality, a much smaller effect of this
interaction is expected, since the strain at the end junctions
tends to keep $r_o-r_i$ large and since the interaction should
depend inversely on $(r_o-r_i)^6$. In any case, the inter-wall
interaction is negligibly small in comparison to values shown in
Fig.~\ref{fig:figure6}(c).
% Note that, however, the above discussion ignores the contribution
% from the non-local inter-wall interactions such as the van der
% Waals interaction, which can be significant if the inter-wall
% separation ($\Delta r = r_o-r_i$) is small. This is because the
% stability analysis based on curvature is completely local. For
% double-walled nanotubes, this can be straightforwardly fixed since
% the non-local contribution can be easily expressed in terms of the
% shape parameters such as $H$ and the radii $r_i$ and $r_o$, and is
% evenly distributed among all constituent atoms. Generically
% speaking, for two-dimensional nanostructures with other geometries
% and topologies the non-local contribution has to be accounted for
% on a case-by-case basis.
In other, non-toroidal structures including helically coiled
CNTs\cite{HCCNT-papers}, the significance of the inter-wall
interaction has to be accounted for on a case-by-case basis.
% one expects that strong van der Waals interaction exists between
% neighboring grooves, whereas the interaction energy is clearly
% unevenly distributed among the atoms.

\subsubsection*{Rotational Symmetry Number Dependence}

To the best of the authors' knowledge, all polygonal carbon
nanotori investigated in the literature had rotational symmetry
numbers $n=5$ or $n=6$. This is largely due to the fact that a
nanotorus can be constructed in a cut-and-paste manner from
graphene\cite{{JCIM},{reference-35-36-a},{stoddarttorus11}} while
preserving its original hexagonal symmetry. Also the majority of
nanotori studied here has been constructed using a scheme
generating isomers with either $D_{nd}$ or $D_{nh}$ point group
symmetry. This constraint reduces greatly the number of possible
isomers, but provides for an easy way to characterize individual
polygonal nanotori by the relative positions of non-hexagonal
rings within a rotational unit cell. With only selected results
for $n=5$ and $n=6$ at hand, no conclusions are possible regarding
the dependence of the stability on the rotational symmetry number.

Such results are presented in Fig.~\ref{fig:figure6}(d), where we
display the curvature energy as a function of the rotational
symmetry number, with $n$ changing from $n=4$ to $n=13$, while all
other shape parameters are fixed. We focus on two structural
nanotori families. The first family, represented by black squares
in Fig.~\ref{fig:figure6}(d), was studied more extensively in the
literature\cite{{reference-21-a},{reference-21-b},{Fowler-PRL},%
{Fe10-T120-PRL}} and is found to be most stable for $n=6$. The
second family, represented by the blue dots, consists of a series
of nanotori with a different distribution of non-hexagonal rings
and is energetically optimal for $n=9$ instead. The specific
shapes of these two families are presented in the Supplemental
Material\cite{torus15-SM}.

Among the nanotori investigated in this study and shown in the
Supplemental material\cite{torus15-SM}, we find that a significant
fraction minimized the curvature energy per atom for a
non-hexagonal rotational symmetry. As a matter of fact, the
distribution of `optimum rotational symmetry numbers $n_{opt}$',
displayed in the Supplemental Material\cite{torus15-SM}, is
roughly a Gaussian centered at $n_{opt}{\approx}7$,
% We find that there is a significant portion of the carbon
% nanotori investigated in this
% study in which the curvature energy per atom is minimized at
% $n{\neq}6$. In fact, the optimal rotational symmetry number is
% roughly Gaussian-distributed centered at $n=7$,
with some extreme outliers at $n_{opt}=12$.
%, see Fig.~S4.
We found that the deviation of $n_{opt}$ from the expected value
$n_{opt}=6$ is an artifact caused by the parametrized Keating
force field, which is penalizing bond length deviations from
$1.42$~{\AA} more than bond angle deviations from $120^\circ$,
while underestimating the penalty for out-of-plane bending and
thus somewhat distorting the optimum geometry.
% This deviation can be attributed to the form of Keating potential,
% which requires the bond lengths to be $1.42$~{\AA} and the bond
% angles to be near $120^\circ$, with more emphasis on the former
% requirement. While these values are met in a perfect hexagonal
% honeycomb, \textit{e.g.} graphene, it is not obvious that the
% sixfold rotational symmetry minimizes the deviation from them when
% a graphitic structure is forced to bend out-of-plane, which is
% always the case for fullerenes and carbon nanotori.

The optimality of the $n=6$ rotational symmetry number can be
restored in the large-torus limit as follows. Let us consider a
cut-and-paste model of a graphitic torus with $n=6$ in the shape
displayed in Fig.~\ref{fig:figure6}(a). It is clear that all
non-hexagonal rings, which determine the angle $\varphi=120^\circ$
between adjacent nanotube segments, will be located in the elbow
joints colored in black, whereas the straight grey-colored
segments will contain only hexagons. In large tori, the relative
role of the elbow joints will play an ever diminishing role, and
strain energy will be determined by the shape of the nanotube
segments. Deviation from $n=6$ will mean that the nanotube
segments need to be bent, which causes extra strain.

% On the other hand, the optimality at $n=6$ can be restored at the
% large torus limit and explained by a simple scaling argument. In
% the cut-and-paste model of nanotori, the curvature energy is
% assumed to be concentrated at the vertices, where the
% non-hexagonal rings are located, and the edges of the model. While
% on the faces of the model the structure of planar graphene
% perfectly fits in, with zero excess energy. At a given rotational
% symmetry number, the number of vertices is constant, the length of
% the edges scales as $\sqrt{N}$, and the area of the faces scales
% as $N$, where $N$ is the total number of atoms. Consequently, the
% excess energy per atom scales as $1/\sqrt{N}$ as acquired from the
% bending energy at the edges. This decays faster to zero in the
% $N\rightarrow\infty$ limit if any dependence on $N$ is introduced
% for the energy of the faces, which is the case if they are allowed
% to bend out-of-plane. This has been confirmed in the fullerene
% case\cite{Siber-paper} and is consistent with our preliminary
% calculations of large nanotorus systems, that for larger polygonal
% nanotori the distribution of optimal $n$ is more narrowly
% distributed around $n=6$.

\section*{Conclusions}%
\label{sec:Conclusions}

We have presented comprehensive analysis on the elastic energy of
carbon nanotori containing either only hexagons (polyhex nanotori)
or also other polygons (polygonal nanotori) on the basis of
continuum elasticity theory. In polyhex nanotori, we found that
depending on the ratio between the major radius $R$ and minor
radius $r$ of the torus, the in-plane and the out-of-plane
contributions to the total elastic energy vary significantly. The
wide CNT rings resembling nanotori, which have been observed
experimentally, display only negligible in-plane strain, whereas
the in-plane strain should exceed the out-of plane strain for
$R/r{\lesssim}20$. In the polygonal nanotori studied here, the
in-plane strain is rather small and the elastic energy obtained
with the continuum method is shown to agree quantitatively with
the results of {\em ab initio} DFT calculations. We also show that
the analytical expression Eq.~(\ref{eqn:dEc_torus}) can serve as a
quick and qualitative reference for the elastic energy of nanotori
once the shape parameters $(R,r)$ are known.

The capability of the current methodology is further demonstrated
by a detailed analysis of the distributions of local excess energy
in individual nanotori. Depending on the relative loci of
non-hexagonal rings, the distribution of Gaussian curvature, mean
curvature, and the local curvature energy can be either localized
or evenly distributed across the nanotorus surface. This analysis
can be extended to other 2D systems with different chemical
composition and shape, and can also be related to the local
stability and chemical reactivity index of different
sites.\cite{{DT238},{SBarraza-Lopez14}}

We have furthermore studied three different sets of polygonal
nanotori with varying shape parameters, including the lateral and
the axial dimension of the nanotori and the rotational symmetry
number. Contrary to the common perception that the most stable
nanotori all have a six-fold symmetry, we find that in smaller
polygonal nanotori, the optimal rotational symmetry number covers
a wide range $4{\lesssim}n_{opt}{\lesssim}12$. Only in the
large-size limit, when the number of non-hexagonal defects is
fixed, $n=6$ emerges as the optimum rotational symmetry number.
Asymptotic analysis on the variation of the other two shape
parameters agrees quantitatively with the numerical results. This
confirms that the current methodology, at least for the systems
investigated here, is applicable across a wide length scale: From
small nanotori where {\em ab initio} calculations are available,
to mesoscopic tori, where continuum elasticity theory applies.
Owing to the broad applicability, we believe that our approach
will provide valuable results pertaining to the thermodynamical
behavior of other large, experimentally observed carbon
nanostructures, where atomic-scale treatment by {\em ab initio}
techniques is not practical.

\begin{acknowledgments}
This work was partly funded by the National Science Foundation
Cooperative Agreement \#EEC-0832785, titled ``NSEC: Center for
High-rate Nanomanufacturing''. Computational resources for this
project were provided by the Michigan State University
High-Performance Computer Center.
\end{acknowledgments}

%%%%%%%%%%%%%%%%%%%%%%%%%%%%%%%%%%%%%%%%%%%%%%%%%%%%%%%%%%%%%%%%%%%%%
%% The appropriate \bibliography command should be placed here.
%% Notice that the class file automatically sets \bibliographystyle
%% and also names the section correctly.
% \bibliographystyle{apsrev4-1}
% \bibliography{torus15}
% \end{document}
%%%%%%%%%%%%%%%%%%%%%%%%%%%%%%%%%%%%%%%%%%%%%%%%%%%%%%%%%%%%%%%%%%%%%

%merlin.mbs apsrev4-1.bst 2010-07-25 4.21a (PWD, AO, DPC) hacked
%Control: key (0)
%Control: author (8) initials jnrlst
%Control: editor formatted (1) identically to author
%Control: production of article title (-1) disabled
%Control: page (0) single
%Control: year (1) truncated
%Control: production of eprint (0) enabled
%

\end{document}